{}

\documentclass[conference]{IEEEtran}

\usepackage{algorithmic}
\usepackage{graphicx}
\usepackage{textcomp}
\usepackage{xcolor}
\usepackage{comment}
\usepackage{fontawesome}
\usepackage{csquotes}
\usepackage{relsize,etoolbox}
\usepackage[rightmargin=0.5pt]{quoting}
\usepackage{booktabs} 

\usepackage{xspace}
\usepackage[normalem]{ulem}
\useunder{\uline}{\ul}{}
\usepackage{multirow}
\usepackage{lscape}
\usepackage{rotating}
\usepackage{supertabular}
\usepackage{longtable}
\usepackage{enumerate}
\usepackage{colortbl}
\usepackage{subcaption}
\usepackage{hhline}
\usepackage{makecell}
\usepackage{caption}
\usepackage{hyperref}
\usepackage{enumitem}
\usepackage{tcolorbox}
\usepackage{tabularx}
\usepackage{lipsum}
\usepackage{balance}
\usepackage{blindtext,graphicx}
\usepackage[absolute]{textpos}
\ifCLASSINFOpdf
\else
\fi
\begin{document}
%
\title{Does Domain Change the Opinion of Individuals on Human Values? A Preliminary Investigation on eHealth Apps End-users}

\author{\IEEEauthorblockN{Humphrey O. Obie\IEEEauthorrefmark{1},
Mojtaba Shahin\IEEEauthorrefmark{1},
John Grundy\IEEEauthorrefmark{1},
Burak Turhan \IEEEauthorrefmark{1} \IEEEauthorrefmark{2},
Li Li\IEEEauthorrefmark{1},
Waqar Hussain\IEEEauthorrefmark{1},
Jon Whittle\IEEEauthorrefmark{3}}
\IEEEauthorblockA{\IEEEauthorrefmark{1}Monash University, Melbourne, Australia}
\IEEEauthorblockA{\IEEEauthorrefmark{2}University of Oulu, Oulu, Finland}
\IEEEauthorblockA{\IEEEauthorrefmark{3}CSIRO's Data61, Melbourne, Australia \\
\{humphrey.obie, mojtaba.shahin, john.grundy, burak.turhan, li.li, waqar.hussain\}@monash.edu, 
jon.whittle@data61.csiro.au}}


\maketitle

\begin{abstract}

The elicitation of end-users’ \textit{human values} -- such as freedom, honesty, transparency, etc -- is important in the development of software systems. 
We carried out two preliminary Q-studies to understand (a) the \textit{general} human value opinion types of eHealth applications (apps) end-users (b) the \textit{eHealth domain} human value opinion types of eHealth apps end-users (c) whether there are differences between the general and eHealth domain opinion types.
Our early results show three value opinion types using \textit{generic} value instruments: (1) fun-loving, success-driven and independent end-user, (2) security-conscious, socially-concerned, and success-driven
end-user, and (3) benevolent, success-driven, and conformist end-user
Our results also show two value opinion types using \textit{domain-specific} value instruments: (1) security-conscious, reputable, and honest end-user, and (2) success-driven, reputable and pain-avoiding end-user.
Given these results, consideration should be given to domain context in the design and application of values elicitation instruments. 

\end{abstract}


%
\IEEEpeerreviewmaketitle

\section{Introduction}
The study of human values and its place in technology has begun to gain interest in the field of software engineering, although research in this area is still in its nascent stage \cite{Obie:2020}. Recent work has covered, amongst other things, value-based requirements engineering -- elicitation of human values in requirements gathering \cite{Thew:2018}; value-sensitive design (VSD) -- a principled manner through which technology can account for values in the design process \cite{Friedman:2013}; and value-sensitive software development (VSSD) -- frameworks for translating values into software features \cite{Aldewereld:2015}.

Software engineering research has mostly concentrated on well-known human values, such as security, privacy and accessibility, with little focus on the broader human values, such as curiosity and honesty \cite{Obie:2020}. However, the recent high profile cases of the violation of human values and their associated consequences in the media have further shown that software systems are not value-agnostic or neutral, e.g. Facebook-Cambridge Analytica scandal \cite{FTC:2020}, Amazon biased same day shipping service \cite{Gralla:2016}. These negative examples have reinforced the need for the full consideration of the human values of relevant stakeholders in the development of software systems.

There has been some preliminary work in the elicitation of stakeholders’ values, using the Schwartz human values model, through both reactive approaches, such as the mining of app reviews \cite{Obie:2020}, and proactive approaches, such as engaging stakeholders through survey instruments \cite{Winter:2018,Shams:2021}. In the proactive elicitation of values, two types of survey instruments have been used: a ``general" human values instrument, e.g., Portrait Values Questionnaire (PVQ) \cite{Shams:2021}, and customised ``domain-specific" instruments, e.g., Values Q-Sort \cite{Winter:2018}. Both types of instruments have been applied in specific technology domains: PVQ has been used to understand the values of end-users for the development of mobile apps in the agriculture domain \cite{Shams:2021} while the Values Q-Sort has been used to understand the values opinion types of software engineers \cite{Winter:2018}. 

Generic instruments such as the PVQ are context-agnostic. This raises the question as to whether they are appropriate for measuring human values in specific domain contexts. Given the probability of a person’s hierarchy of values varying depending on the contextual domain they are in it also raises the question of whether the human values measured using generic instruments in a specific domain are comparable to those measured using a domain-specific instrument customised for the domain. For example, Winter et al. \cite{Winter:2019} notes, \textit{“applying Schwartz’s values to a specific sector of life may put to test a relational values model designed to apply to life generally”}. Hence it is important to explore the degree to which generic instruments are effective in specific domains and whether customised instruments are better suited to those domains.

To understand the differences between the application of a generic human values instrument and a domain-specific customised human values instrument, we conducted a preliminary investigation, following the Q-methodology with 8 participants. Our preliminary investigation consisted of two Q-studies: Q-Study 1 was conducted with generic Q-statements from the PVQ and Q-study 2 was conducted with customised Q-statements developed from the eHealth apps domain. \textbf{We chose the eHealth domain as it is a combination of health -- a basic human need and modern technology \cite{Akerkar:2004} -- with the potential to influence a person’s value hierarchy,} e.g., a person who generally ranks autonomy of choice highly may lower that value in the health domain when faced with an important decision and defer to an expert opinion. \textbf{We considered the Q-methodology particularly appropriate for our preliminary investigation because it is well suited to show the \textit{“inter-subjective orderings of beliefs that are shared among people”} \cite{Webler:2009}}. Also, in the study of human values, \textbf{the Q-methodology has been shown to reflect the \textit{“relational nature of values by asking participants to consider statements together and make trade-offs”}} \cite{Winter:2019}.
Our preliminary investigation seeks to answer the following  research questions:

\textit{\textbf{RQ1 What are the general human value opinion types of eHealth apps end-users?} }

\textit{\textbf{RQ2 What are the eHealth-domain human value opinion types of eHealth apps end-users?}}

\textit{\textbf{RQ3 Are there differences in the elicited human values opinion types based on the application of a generic values instrument versus a context-specific values instrument?}}


\section{Background and Related Work}
\subsection{Theoretical Frameworks for Human Values}

Most studies on human values are based on the theories developed in the social sciences \cite{Whittle:2021}; specifically, the discipline of social psychology has provided rich insight into how values are developed and propagated within social groups \cite{Winter:2019}.

Rokeach presented human values as determiners of behaviour and attitude \cite{Rokeach:1973}. The Rokeach value scale categorises 36 human values into 2 main categories: 18 terminal values which describe goals in life and 18 instrumental values which describe modes of conduct.
The fundamental theory of human values posits human values as a guide for actions and a vehicle for expressing need \cite{Gouveia:2014}. However, the most widely accepted and adopted theory of human values is the Schwartz theory of basic human values \cite{Schwartz:1992}; it has seen adoption in other fields beyond the social sciences, e.g., software engineering \cite{Winter:2018,Obie:2020}.

The Schwartz theory of basic human values is based on survey studies conducted in several countries covering various dimensions including age, gender, cultural practices, and geography. The Schwartz theory categorises 58 human values into 10 categories that are “structured in similar ways across culturally diverse groups.” \cite{Schwartz:1992}. While this theory has been well applied more generally resulting in the discovery of general human value types, we aim to explore values in a particular technology domain -- eHealth. 




\begin{table}
\scriptsize
\centering
\caption{Value categories and descriptions \cite{Schwartz:1992}}
\label{tab:values}
\begin{tabular}{lp{6cm}} 
\hline
\textbf{Value Category} & \textbf{Description (motivational goals) }                                                                                         \\ 
\hline
Self-direction & Independent thought and action - choosing, creating, exploring                                                            \\ 
\hline
Stimulation    & Excitement, novelty, and challenge in life                                                                                \\ 
\hline
Hedonism       & Pleasure or sensuous gratification for oneself                                                                            \\ 
\hline
Achievement    & Personal success through demonstrating competence according to social standards                                           \\ 
\hline
Power          & Social status and prestige, control or dominance over people and resources                                                \\ 
\hline
Security~      & Safety, harmony, and stability of society, of relationships, and of self                                                  \\ 
\hline
Conformity     & Restraint of actions, inclinations, and impulses likely to upset or harm others and violate social expectations or norms  \\ 
\hline
Tradition      & Respect, commitment, and acceptance of the customs and ideas that one's culture or religion provides                      \\ 
\hline
Benevolence    & Preserving and enhancing the welfare of those with whom one is in frequent personal contact                               \\ 
\hline
Universalism   & Understanding, appreciation, tolerance, and protection for the welfare of all people and for nature                       \\
\hline
\end{tabular}
\end{table}



\subsection{Measuring Human Values in Software}
Recent studies have sought to understand the reflection and violation of human values in software applications \cite{Shams:2020,Obie:2020}. Focusing on the agriculture domain, Shams et al. manually analysed 1,522 user app reviews from 29 Bangladeshi agricultural mobile apps, showing the missing values and those desired by Bangladeshi farmers in mobile apps \cite{Shams:2020}. Similarly, to understand the violation of human values in popular mobile apps in varied categories, Obie et al. analysed 22,119 app reviews using natural language processing techniques based on a values dictionary built upon Schwartz theory \cite{Obie:2020}. Both studies do not directly measure the values of users or other stakeholders but instead utilise user app reviews as a proxy for eliciting their values (and violations of their values by apps).

Some studies have directly elicited and measured the values of stakeholders, e.g., end-users, software developers, using various instruments, albeit based on Schwartz theory. Shams et al elicited the values of end-users in the agriculture domain, specifically female farmers in Bangladesh \cite{Shams:2021}. They adopted the well-known  PVQ instrument to measure the end-users’ human values. 
However, the PVQ is generic and is not tailored to any particular domain context. Moreover, taking a slightly different approach and using a different instrument, Winter et al. measured the values of software engineers (software engineering domain) \cite{Winter:2018}. Winter et al. developed the \textit{Values Q-Sort} to fit the domain of software engineering. While based on the Schwartz theory, the \textit{Values Q-Sort} “has been designed so that the chosen statements are both related to an appropriate model of human values and relevant to the community being studied (the SE community).”

\begin{table*}

\caption{Q-statements}
\label{tab:statements}
\resizebox{\textwidth}{!}{%
{\renewcommand{\arraystretch}{1.2}
\begin{tabular}{lll} 
\hline
\textbf{ID} & \textbf{General Values Statements}                                                                                                                                 & \textbf{Value Category}  \\ 
\hline
GS1         & It is important to me to make my own decisions about what I do. I like to be free to plan and to choose my activities for myself.                                  & Self-direction  \\
GS2         & I like surprises. It is important to me to have an exciting life.                                                                                                  & Stimulation     \\
GS3         & It's very important to me to show my abilities. I want people to admire what I do.                                                                                 & Achievement     \\
GS4         & I believe it is best to do things in traditional ways. It is important to me to follow the customs I have learned.                                                 & Tradition       \\
GS5         & It's very important to me to help the people around me. I want to care for other people.                                                                           & Benevolence     \\
GS6         & I always want to be the one who makes the decisions. I like to be the leader.                                                                                      & Power           \\
GS7         & I believe that people should do what they are told. I think people should follow rules at all times, even when no-one is watching.                                 & Conformity      \\
GS8         & I think it is important that every person in the world be treated equally. I want justice for everybody, even for people I donít know.                             & Universalism    \\
GS9         & I really want to enjoy life. Having a good time is very important to me.                                                                                           & Hedonism        \\
GS10        & It is important to me to live in secure surroundings. I avoid anything that might endanger my safety.                                                              & Security        \\ 
\hline
\textbf{ID} & \textbf{eHealth Values Statements}                                                                                                                                 &  \textbf{Value Category}               \\ 
\hline
ES1         & The functionalities of this hearing aid app are not accessible without knowing my location.                                                                        & Self-direction  \\
ES2         & The new dashboard of this health app has taken a lot of fun out of the experience. It's terrible and consists of dull circles and colours. It's boring.            & Stimulation     \\
ES3         & I've spent most of the day trying to print out my daughter's immunisation record generated by the app, but to no avail. It's very frustrating.                     & Achievement     \\
ES4         & This food diary app is trying to force me to go premium. I would rather stick with my physical food diary; at least my pen and paper are reliable and won't crash. & Tradition       \\
ES5         & This eHealth app sold my data to ambulance chasers, so I requested my account be deleted. Nine months later my account is still active. They lied to me.           & Benevolence     \\
ES6         & Now I get rude marketers calling after installing this health app. You don't care who uses your services which in turn tarnishes your product. I won't refer you.  & Power           \\
ES7         & I'm forced to use this health app to make an online claim. I'm required to take a picture of the receipt through the app even though the app quality is terrible.  & Conformity      \\
ES8         & This healthy food app only caters for folks in the USA, as it only uses the imperial system. Customers from other countries find it difficult to use.              & Universalism    \\
ES9         & Before I used this eHealth app, fear of pain never entered my mind. Now I have to worry about extra fear and anxiety that wasn't there before.                     & Hedonism        \\
ES10        & This eHealth app shows numerous defibrillator locations but many are not available for public use. This misleading information could lead to a loss of life.       & Security        \\
\hline
\end{tabular}
}\quad
}

\end{table*}

Other studies have also reflected on the importance of the contextual domain (even within a broader domain like software engineering) in the coverage of human values. Hussain et al. \cite{Hussain:2020} observe that the  \textit{“coverage of values in a given software has to be a contextual decision where addressing more values does not necessarily mean a better software.”} The importance of  domain considerations and the tailoring of value instruments to fit specific contextual domains instead of the wholesale application of a generic instrument like the PVQ may be more important than previously considered. 


\section{Q-Methodology}
To answer our research questions, we used Q-methodology proposed by  Stephenson \cite{stephenson1935technique}. Q-methodology is a technique to uncover patterns in diverse opinions, beliefs, concerns, or attitudes of individuals on a topic \cite{Webler:2009, watts2012doing}. The diversity of the opinions can be elicited by Q-methodology, although they might be prevalent within a population \cite{Webler:2009}. Q-methodology achieves this by leveraging the benefits of both qualitative and quantitative research approaches. Unlike more common social research methods (e.g. use of surveys), Q-methodology provides a qualitative explanation and comparison of the entire opinions of participants, and its quantitative characteristic detects the more nuanced differences between opinions \cite{Webler:2009}. Another benefit of Q-methodology is that it does not require a large sample of participants \cite{Webler:2009}. 
\subsection{Statements Development}
The first step in Q-methodology is to develop a concourse of statements on the topic of interest \cite{watts2012doing, Webler:2009}. Statements can be retrieved from different resources such as interviews, literature review, etc \cite{Webler:2009}. 
Our goal is to understand the possible difference between the opinion of eHealth apps end-users on general human values (Q-study 1) and their opinion on the eHealth domain human values (Q-study 2). 

First, we used PVQ to develop the concourse of statements for Q-study 1. The PVQ is the most widely used human values instrument \cite{Shams:2021} and includes 40 statements, which measure 10 human value categories (See Table \ref{tab:values}). Second, we developed 40 eHealth domain human values statements based on 40 eHealth apps user reviews for Q-study 2 to match the number of the statements in PVQ. We then conducted two rounds of a pilot study with 2 persons each on the 40 eHealth values statements. The feedback from the pilot study and insight from previous research \cite{Winter:2019} showing too many questions as cognitively overwhelming and time consuming enabled us to identify ambiguous and redundant eHealth values statements. We reduced the 40 eHealth values statements to 10 statements - one statement for each PVQ value category (See Table \ref{tab:statements}). Next, to compare Q-study 1 and Q-study 2, we had to reduce the PVQ 40 statements to 10 statements. To this end, we chose only one statement from each human value category relevant to the eHealth statement in that value category. Thus participants had to rank 20 statements altogether instead of 80 statements, so as not to cognitively overwhelm them \cite{Winter:2019}.

\subsection{Participants Recruitment}
\begin{table}
\caption{Participants' Demographics}
\label{tab:demographics}
\scriptsize
{\renewcommand{\arraystretch}{1.2}
\begin{tabular}{lp{6cm}}
\hline
\textbf{Characteristic}                  & \textbf{Number of Participants (n = 8)     }                                                                                              \\ \hline
Gender                          & Female (4), Male (4)                                                                                                             \\
Cultural background                          & Africa (2), Middle East (2), Asian (2), Australia (2)                                                                            \\
Age                             & 26-35 (6), 36-45 (2)                                                                             \\
Education level                 & Bachelor's degree (3), Master's degree (3), PhD degree (2) \\
eHealth apps usage  & Daily (3), 2-3 days a week (2), 4 - 5 days a week (1), Once a week (1), Once a month (1)                        \\ \hline
\end{tabular}%
}\quad
\end{table}
Q-methodology does not need a large number of participants and participants do not need to be representative of the population \cite{watts2012doing}. Hence, it is common to use a purposive sampling method to recruit participants. We purposively selected participants with different characteristics, cultures, genders, and nationalities to collect diverse and well-informed opinions \cite{webler2009using}. Besides this, participants had to use eHealth apps. We reached out to our personal contacts, who we thought use eHealth apps, via email and social media such as WhatsApp. We asked them if they use any eHealth app. Nine persons indicated that they use such apps. We chose eight participants for a balance and also because we had only ten statements. Q-methodology requires the number of sample to be less than the number of Q-statements \cite{watts2012doing, Winter:2019}. Finally, the selected eight participants agreed and completed our study. Table \ref{tab:demographics} shows an overview of the participants' demographics.

\subsection{Data Collection}
We used an interactive web application\footnote{https://github.com/shawnbanasick/easy-htmlq}  to ask the participants to rank the 10 statements in each Q-study. The process of data collection included three steps. In \textbf{Step 1}, the participants were asked to read the statements and split them into three piles. One pile was for the statements that were ranked most important by the participants (the “Most Important” pile). A “Least Important” pile was for the statements they considered least important, and the final pile was for the rest (the “Neutral” pile). In \textbf{Step 2}, we asked the participants to read the statements again in the three piles and place them in the Q-sort grid. For example, the participants had to read the statements from the “Most Important” pile again and select the statement they consider most important and place it on the right side of the Q-sort grid below the “+2”. In \textbf{Step 3}, the two statements that the participants selected as most important and least important were shown to the participants to seek the participants’ motivations for their ranking of these two statements. After participants ranked the 20 statements, they filled a short questionnaire, which collected their demographic information. The replication package is available\footnote{https://doi.org/10.5281/zenodo.5105639}. 

\subsection{Data Analysis}
We used a Q-methodology application\footnote{https://shawnbanasick.github.io/ken-q-analysis-beta}   to uncover the opinions of eHealth apps end-users on general human values statements and eHealth domain human values statements.
We input the Q-sort data from both Q-studies into the app. It supports correlation and by-person factor analysis and varimax rotation to discover the factors that constitute clusters of participants with similar opinions. We carried out this analysis on the Q-sort data from the two studies; the Q-Sort data based on the PVQ and Q-sort data based on the eHealth apps domain. The factors extracted, using factor centroid analysis show statistically significant patterns in opinions of the participants. A factor is assigned an Eigenvalue -  the sum of the square of the individual Q-sort loadings onto the factor \cite{Winter:2018}. The higher the Eigenvalue, the more variance explained by the factor. Although, typically Q-methodology can extract up to 8 factors, only 3 to 4 factors have any real value \cite{watts2012doing}.


For our analysis, we chose 3 factors in Q-study-1 and 2 factors in Q-study-2. The chosen factors met the following criteria: the factors must have a minimum Eigenvalue of 0.85 and contain distinguishing statements. Each of these factors constitute an opinion type or viewpoint of an end-user.

\subsection{Threats to Validity}

\textit{Population sample: }
The number of participants for this investigation is 8 -  a relatively small number. However, our participants selection was governed by the principles of Q-methodology; the number of participants should be fewer than the number of Q-statements \cite{watts2012doing}. For example, the work by Winter et al. \cite{Winter:2019} had 19 Q-statements and 12 participants. Our recruitment was done after finalising the 10 Q-statements for each Q-study, hence we had 8 participants to keep with the tenets of Q-methodology. 
Moreover, this paper reports early results from our preliminary investigation.

\textit{Generalisability: }
Like most Q-studies, ours is exploratory in nature. Because of our application of a purposive sampling instead of random sampling technique, generalisations may not be made beyond the cohort of participants. However, the usefulness of Q-methodology comes from being able to uncover clusters of opinions \cite{Winter:2018}. And once these clusters have been identified, subsequent testing can be conducted on larger samples using standard variance analytic methods. 



\section{Emerging Results and Discussion}
\label{sec:results}
Below we report the emerging results from our preliminary investigation. We assign a label (based on the results) to each factor. The labels serve as a shorthand identification and description of what the factors are about \cite{Webler:2009}.

\subsection{Q-study - 1: Human Values Based on a General Values Instrument (RQ1)}
\label{sec:Qgeneral}

EHealth app end-users appear to hold the following three opinion types based on extracted factors:

\subsubsection{The fun-loving, success-driven and independent end-user:}
This factor explains 44\% of the variance in this study. The highest rated statement for this end-user opinion type is that “I really want to enjoy life. Having a good time is very important to me.” This statement is significant for $p<0.01$ and is ranked higher in this factor than in all of the other factors based on its z-score. When prodded for why they highly ranked this statement, a participant responded, \textit{“I think being intentional about having a good time in life is important for a balanced life. And I am always happier when I remember the good times I have had already.”} Furthermore, this end-user opinion type values success and wants their achievements to be admired by others. They are also driven by the need to exercise autonomy in the choices and decisions they make.

\subsubsection{The security-conscious, socially-concerned, and success-driven end-user:}
This factor explains 18\% of the variance obtained in this study. This end-user opinion type prioritises security and avoids anything that might endanger their safety. The Q-statement corresponding to security is significant for $p<0.01$ and is higher in this factor than in other factors based on its z-score. A participant commented concerning their ranking, \textit{“I think it is the most basic requirement for a comfortable life to have a safe environment”}. Also, while this end-user opinion values equality and wants everyone to be treated equally, they also want to be admired for their success. In addition, of low priority to this end-user opinion type is the value of benevolence; the associated benevolence Q-statement ranks lower in this factor than in other factors and is significant for $p<0.05$.

\subsubsection{The benevolent, success-driven, and conformist end-user:}
This factor explains 11\% of the variance in this study. This end-user opinion type highly values benevolence and wants to help the people around them. The associated Q-statement for benevolence is significant for $p<0.01$ and ranks higher in this factor than in other factors. A participant captured their ranking thus, \textit{“People are important, and it is important to care for them”}. Also, this end-user type also believes that people should conform to laid down rules at all times. Additionally, they also value the admiration that comes from being successful.

\subsection{Q-study - 2: Human Values Based on an eHealth Domain Values Intrument (RQ2)}
\label{sec:Qehealth}


The end-users hold the following two opinion types based on the extracted factors:

\subsubsection{The security-conscious, reputable, and honest end-user:}
This factor explains 31\% of the variance obtained in this study. The statement rated highest by this end-user opinion type concerns the safety of health apps and the security of lives; this statement is significant for $p<0.01$ and ranks higher in this factor than in all other factors. With respect to safety and security, a participant remarked, \textit{“I would not use an app that had unreliable information especially when it’s a matter of life and death”}. Furthermore, this end-user opinion type values reputation the exertion of their power. They also value honesty in software providers in dealing with their customers.

\subsubsection{The success-driven, reputable and pain-avoiding end-user:}
This factor contributes 16\% of the variance in this study. This end-user opinion type values, first and foremost, being able to achieve a desired goal. An example comment from a participant captures this: \textit{“If I can’t easily access data from the app, then what’s the point?”} This end-user opinion type also cares about reputation and would exert their power when needed. They also highly rank the avoidance of pain and negative feelings when it comes to health apps. However, this opinion type places low value on honesty from software providers. The lowly ranked associated statement with the value of honesty is significant for $p<0.01$ and is ranked lower in this factor than in all other factors.

\subsection{Differences in Human Values Opinion Types Based on Instrument Type (RQ3)}

Our results show that there are differences in the human value opinion types of end-users when different instruments are applied. 
Of the three human value opinion types from the general Q-study and two human value opinion types from the eHealth Q-study, the only commonalities are \textit{security-conscious} and \textit{success-driven} -- everything else is unique to their respective categories.
It is interesting to note that the same end-users, when placed in a different domain context (eHealth Q-study) rated values differently than in the general domain (general Q-study). In some cases, end-users chose completely different values while in other cases a change in the hierarchy of values.

\section{Implications and Conclusion}
Although this study is a preliminary investigation, our results show that \textbf{the hierarchy of values may well vary depending on the context of the end-users’ domain or experience}, and this {may be different from their general value hierarchy}. It is probable that \textbf{end-users encounter different human value trade-offs} as they navigate through different domain contexts, without necessarily undermining their personal general human values. For example, an end-user who is generally a non-conformist may relegate their value of autonomy within the eHealth domain. Our preliminary results suggest \textbf{the need to develop customised values instruments} when eliciting human values in specific domain contexts. This is instead of relying on wholesale application of generic instruments that \textbf{may not effectively probe human values in specific domain contexts} because the instruments themselves lack contextual significance. 

In future, we plan on extending this early work by conducting interviews to obtain qualitative data from larger samples to dive deeper into individuals' interpretation of values in eHealth and other domain-specific contexts. Conducting group surveys with larger samples and analysis with standard variance analytic methods is another plan.

\section*{Acknowledgements}
\footnotesize
This work is supported by ARC Discovery Grant DP200100020. Grundy is supported by ARC Laureate Fellowship FL190100035. Li is supported by ARC DECRA DE200100016.

\bibliographystyle{IEEEtran}

\bibliography{myref}

\end{document}